\documentclass[12pt]{article}
\let\section=\subsection     \let\subsection=\subsubsection                
\usepackage{graphicx}

\begin{document}
\begin{center}
   {\large \bf Pion Interferometry:}
   {\large \bf Recent results from SPS}\\[5mm]
   Harald~Appelsh\"auser \\[5mm]
   {\small \it  Physikalisches Institut der Universit\"at Heidelberg \\
   Philosophenweg 12, D-69120 Heidelberg, Germany \\[8mm] }
\end{center}

\begin{abstract}\noindent
Recent HBT results from the CERES experiment at the
SPS are reviewed. Emphasis is put on the 
centrality and beam energy dependence, and the results are put into perspective 
with results at lower and higher beam energies. The rather weak beam energy
dependence of the HBT radii may be understood in terms of a transition from baryon to 
pion dominated freeze-out. The observed short lifetimes and emission durations
are presently in contradiction to results from model calculations. 
\end{abstract}

\section{Introduction}

While single particle momentum distributions give only indirect insight
to the space-time evolution of the system created in high energy
heavy ion collisions, the lifetime and the spatial extent of the
pion source as well as the existence of collective velocity fields at the time
of thermal freezeout can be probed by the study of Bose-Einstein momentum  
correlations of identical pions via HBT interferometry~\cite{pratt1}. 
The width of
the correlation peak at vanishing relative momenta reflects the
so-called length of homogeneity of the pion emitting source. 
Only in static sources can the length of homogeneity, in the following
also called 'source radius', be interpreted
as the true geometrical size of the system.
In a dynamic system, the occurence of space-momentum correlations of the 
emitted particles due to collective expansion generally leads to
a reduction of the observed source radii, depending on
the strength of the expansion and the thermal velocity of the
pions $\sqrt{T_{f}/m_{t}}$ at thermal freeze-out~\cite{maksin}. 
A differential analysis of the HBT correlations in bins
of the pair transverse momentum 
$k_{t}$=$\frac{1}{2}|\vec{p_{t,1}}+\vec{p_{t,2}}|$ 
thus provides valuable information about the properties of the collective
expansion of the system~\cite{prattcso}.

To obtain most detailed information about the space-time evolution
the three-momentum difference vector $\vec{q}$ of two like-sign pions
is decomposed 
into components, $\vec{q}$=$(q_{\rm long},q_{\rm side},q_{\rm out})$. 
Following Bertsch and Pratt~\cite{berpra}, $q_{\rm long}$ is the momentum difference along
the beam direction, calculated in the longitudinal rest frame (LCMS) of the pair,
$q_{\rm out}$ is parallel to the pair transverse momentum $\vec{k_{t}}$ and 
$q_{\rm side}$ is perpendicular to $q_{\rm long}$ and $q_{\rm out}$.

The correlation function is defined as the ratio 
$C_{2}(\vec{q})$=$A_{2}(\vec{q})/B_{2}(\vec{q})$
where the 'signal' $A_{2}(\vec{q})$ is the probability to find a pair with momentum difference
$\vec{q}$ in a given event 
and the 'background' $B_{2}(\vec{q})$ is the corresponding mixed-event distribution.

The normalized correlation functions are fit by a Gaussian: 
\begin{eqnarray}
   C_{2}(\vec{q})&=& 1+\lambda \exp(-R_{\rm long}^{2}q_{\rm long}^{2}
                                    -R_{\rm side}^{2}q_{\rm side}^{2} \nonumber \\
                 &&                   -R_{\rm out}^{2}q_{\rm out}^{2}
                                    -2R_{\rm outlong}^{2}q_{\rm out}q_{\rm long}),
\end{eqnarray}

with $R_{\rm long}$, $R_{\rm side}$, $R_{\rm out}$ being the Gaussian source
radii and $\lambda$ the correlation strength. 
The cross-term $R_{\rm outlong}^{2}$ appears as a consequence of space-time
correlations in non-boost-invariant systems.

\section{Results from SPS}
Pion HBT interferometry analyses have been performed by a number of experiments
at SPS ~\cite{na49hbt,na44hbt,wa98hbt,wa97hbt}. 
A systematic study of the centrality dependence of HBT radii around
midrapidity from Pb+Au collisions at all
three presently available SPS energies, 40, 80, and 158 AGeV, was recently
presented by the CERES collaboration~\cite{na45hbt}. Details of their analysis can be found
in~\cite{heinz}.

\vspace{-0.5cm}

\begin{center}
   \includegraphics[width=15cm]{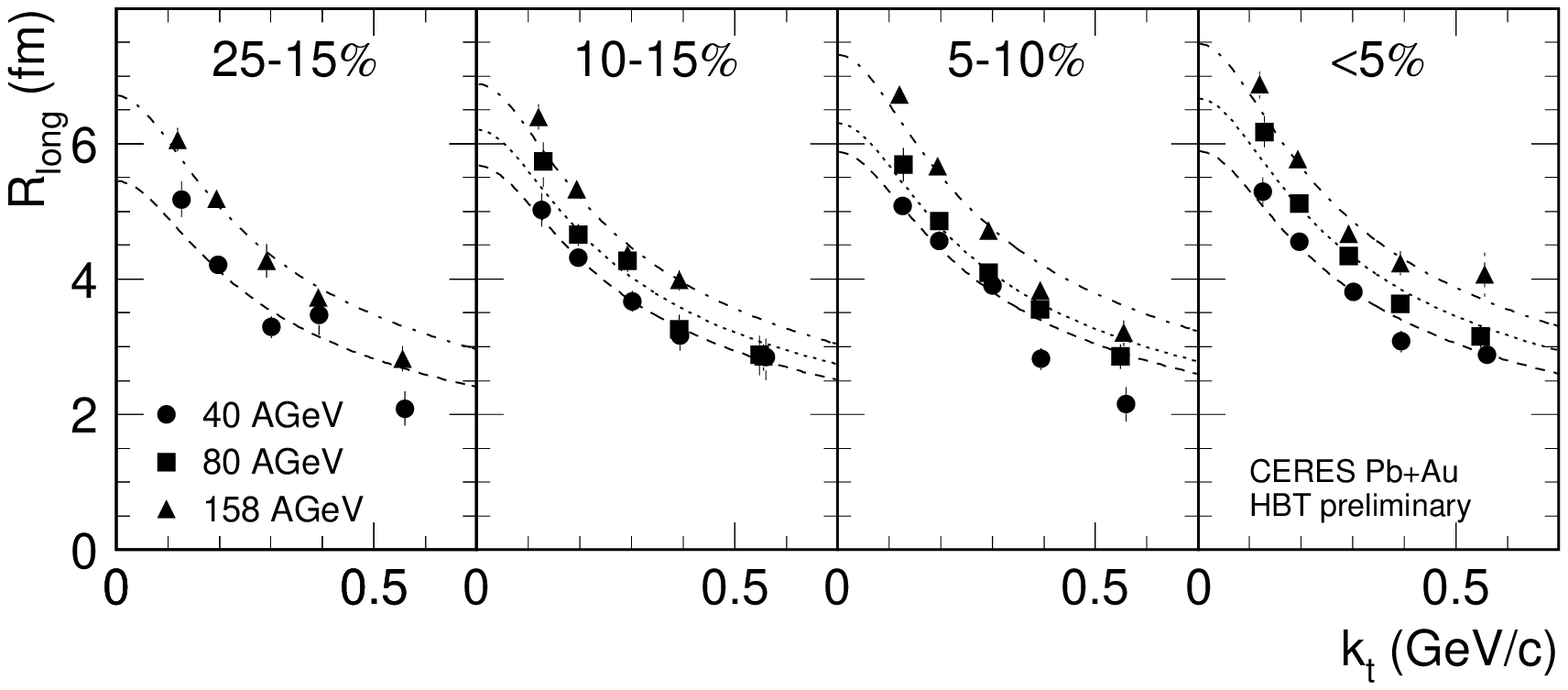}\\
   \parbox{14cm}
        {\centerline {\footnotesize 
        Fig.~1: $R_{\rm long}$ vs.~$k_{t}$~\cite{heinz}. The lines represent fits to the data (see text).}}
\end{center}

\vspace{-0.2cm}

The $k_{t}$-dependence of the longitudinal source parameter $R_{\rm long}$ 
is shown in Fig.~1 for the three different beam energies and as 
function of the centrality of the collision. Similar to previous studies, 
a strong decrease of $R_{\rm long}$ with $k_{t}$ is observed, characteristic
for a strong collective longitudinal expansion, where the length of 
homogeneity is entirely saturated by the thermal length scale $\sim\sqrt{T_{f}/m_{t}}$.
For the case of longitudinal boost-invariance,
$R_{\rm long}$ can be connected to the lifetime $\tau_{s}$ of the system,
the time elapsed between the onset of expansion and kinetic freeze-out by
$R_{\rm long}$=$\tau_{s}(T_{f}/m_{t})^{\frac{1}{2}}$~\cite{maksin}. 

The slight but systematic overall increase of $R_{\rm long}$ with
centrality and beam energie is reflected in a correspondingly 
longer lifetime when applying the above expression to the
data and assuming $T_{f}$=120~MeV (see fits in Fig.~1). The observed lifetimes
at the SPS are 6-8~fm/c.

\vspace{-0.5cm}

\begin{center}
   \includegraphics[width=15cm]{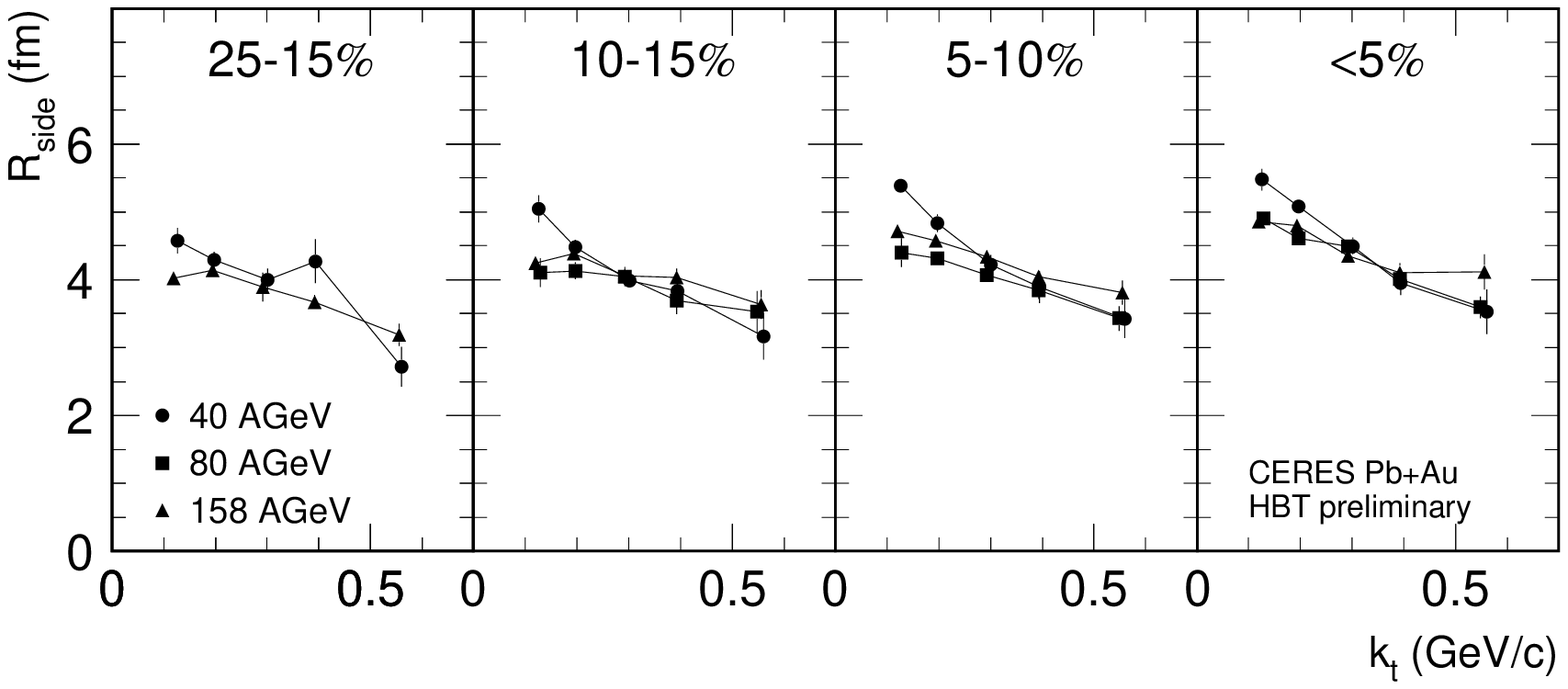}\\
   \parbox{14cm}
        {\centerline {\footnotesize 
        Fig.~2: $R_{\rm side}$ vs.~$k_{t}$~\cite{heinz}. }}
\end{center}

\vspace{-0.5cm}
\vspace{-0.5cm}

\begin{center}
   \includegraphics[width=15cm]{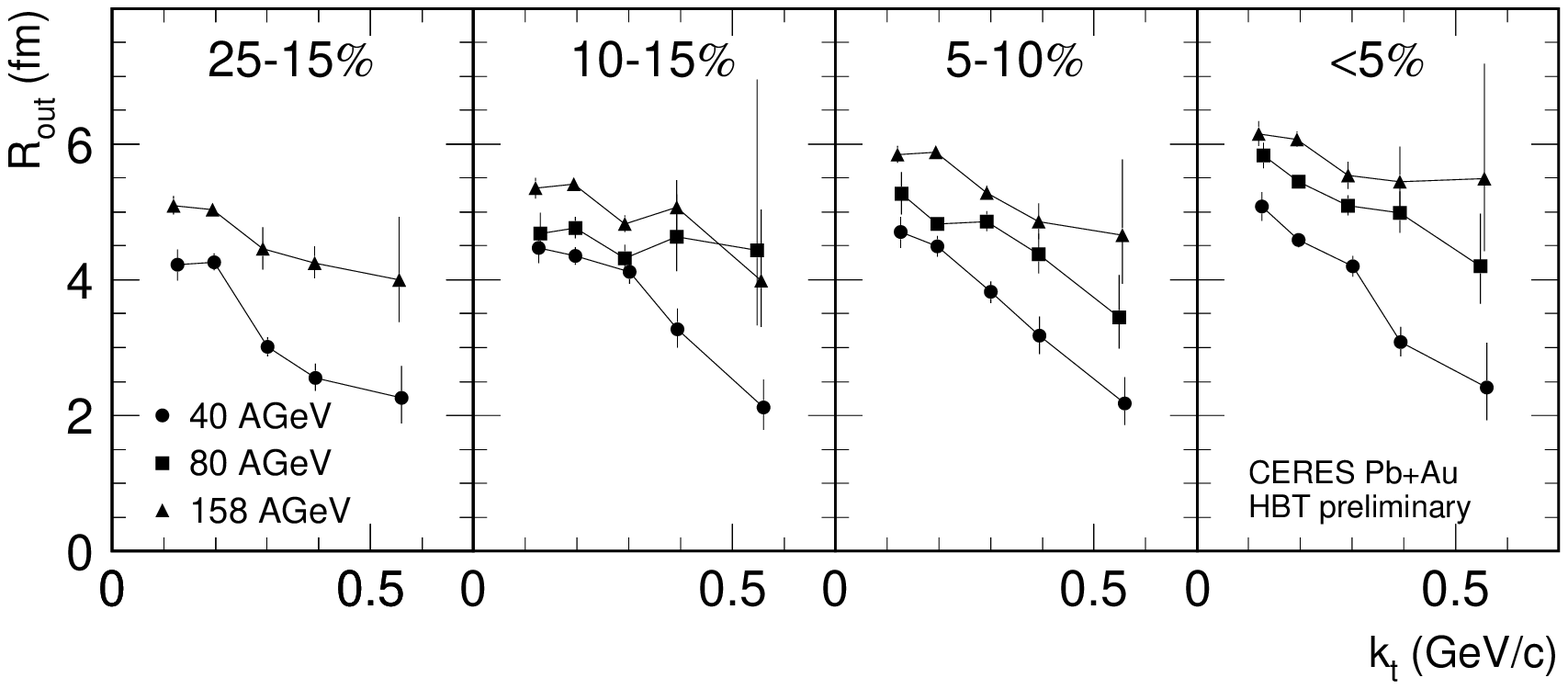}\\
   \parbox{14cm}
        {\centerline {\footnotesize 
        Fig.~3: $R_{\rm out}$ vs.~$k_{t}$~\cite{heinz}.}}
	\label{fig:rout}
\end{center}

\vspace{-0.5cm}

The transverse source parameter $R_{\rm side}$ is closely related to the
{\em true} geometrical transverse extension of the system at freeze-out.
Consequently, a weaker $k_{t}$-dependence is observed for $R_{\rm side}$
(Fig.~2), indicating the presence of radial flow. A slight but systematic increase 
of $R_{\rm side}$
with centrality is measured at all beam energies, as expected from simple collision
geometry. The beam energy dependence is surprising: largest $R_{\rm side}$ and the
strongest $k_{t}$-dependence are observed at the lowest energy. This led 
previously to 
the conclusion that radial flow may reach a maximum at the lower SPS energies~\cite{ceresqm}.
We will discuss below that the extraction of the radial flow velocity 
from the $k_{t}$-dependence of $R_{\rm side}$ at the
lower SPS energies might be questionable. At 158 AGeV, a radial flow velocity
of about 0.5$c$ was obtained from the analysis of single particle
$m_{t}$-spectra and 
$R_{\rm side}(k_{t})$~\cite{na49hbt,tomwiedheinz}.

Many authors have discussed that a strong first order phase transition
with a large latent heat
can lead to a retardation of hadronization during the mixed phase 
and consequently to a long duration of pion
emission~\cite{berpra,bergong,berbrown,rischke,rigyu}. 
This should be observable via a strong increase of the
outward source parameter $R_{\rm out}$ with respect to $R_{\rm side}$. 
However, no long-lived source has been observed so far. At SPS, $R_{\rm out}$
(Fig.~3)
is similar to $R_{\rm side}$, indicating a short pion emission duration
and a sudden freeze-out.

\section{Beam energy dependence}
A large amount of pion interferometry data have been published by experiments
at AGS, SPS, and RHIC. This allows for a systematic study of the source parameters
over a wide range of beam energies. It has been argued that the 
{\em HBT null effect}~\cite{gyulint},
the absence of any 
beam energy dependence, in particular when going to RHIC, is the actual 
surprising result from HBT interferometry.

In Fig.~4 are shown the $k_{t}$-dependences of $R_{\rm long}$,$R_{\rm side}$,
and $R_{\rm out}$ in central Pb(Aa)+Pb(Au) collisions 
at different beam energies~\cite{e895hbt,heinz,starhbt}.
Indeed, no dramatic variation of any of the source parameters can be observed.
However, a closer inspection reveals some interesting features. $R_{\rm long}$
is approximately constant from AGS to the lower SPS energies, but develops
a significant increase within the SPS regime and towards RHIC, indicating a smooth
increase of the lifetime.

Most interesting is the behaviour of $R_{\rm side}$: It is gradually {\em decreasing}
at small $k_{t}$ up to top SPS energy, connected with a continuous {\em flattening}
of the $k_{t}$-dependence. At RHIC, $R_{\rm side}$ is again larger than at SPS
while the shape is not yet well measured, at least the $\pi^{-}\pi^{-}$ sample looks
rather flat.
Naivly, the flattening indicates a {\em decrease} of the radial flow velocity
as function of beam energy, in contradiction to the present 
understanding~\cite{xuqm}.

$R_{\rm out}$ shows a rather weak energy dependence,
possibly indicating a shallow minimum at the lower SPS energy.

\vspace{-0.5cm}
 
\begin{center}
   \includegraphics[width=15cm]{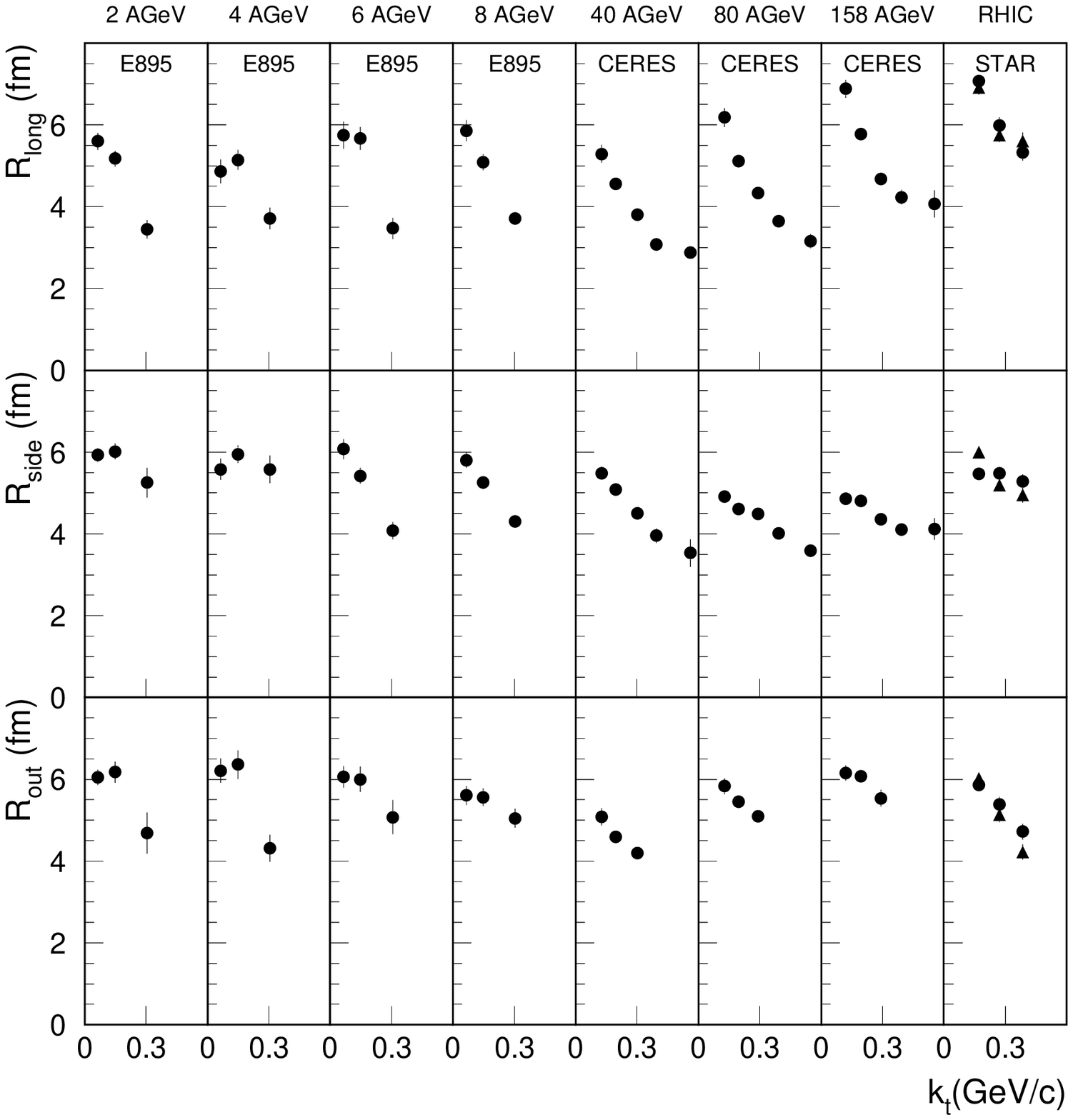}\\
   \parbox{14cm}
        {\footnotesize 
        Fig.~4: $k_{t}$-dependence of HBT radii around midrapidity in central Pb(Au)+Pb(Au) 
	collisions at different beam energies.
	The data are from \cite{e895hbt,heinz,starhbt}.}
	\label{fig:edep}
\end{center}

\vspace{-0.5cm}

A straight-forward investigation of the freeze-out properties can
be performed by relating the measured source parameters to an effective
freeze-out volume, $V_{f}$=$2\pi R_{\rm long}R_{\rm side}^{2}$.
Assuming freeze-out at constant density~\cite{pom}, 
we expect $V_{f}$ to scale linearly
with the charged particle multiplicity. Fig.~5 shows 
$V_{f}$ as function of the number of participants at 40, 80, and 158 AGeV.
A linear scaling with $N_{\rm part}$ is indeed observed at all three energies, consistent
with the assumption of a constant freeze-out density, since the number of
charged particles was found to scale close to linear with $N_{\rm part}$ at 
SPS~\cite{wa98mult}.

\vspace{-0.5cm}

\begin{center}
   \includegraphics[width=15cm]{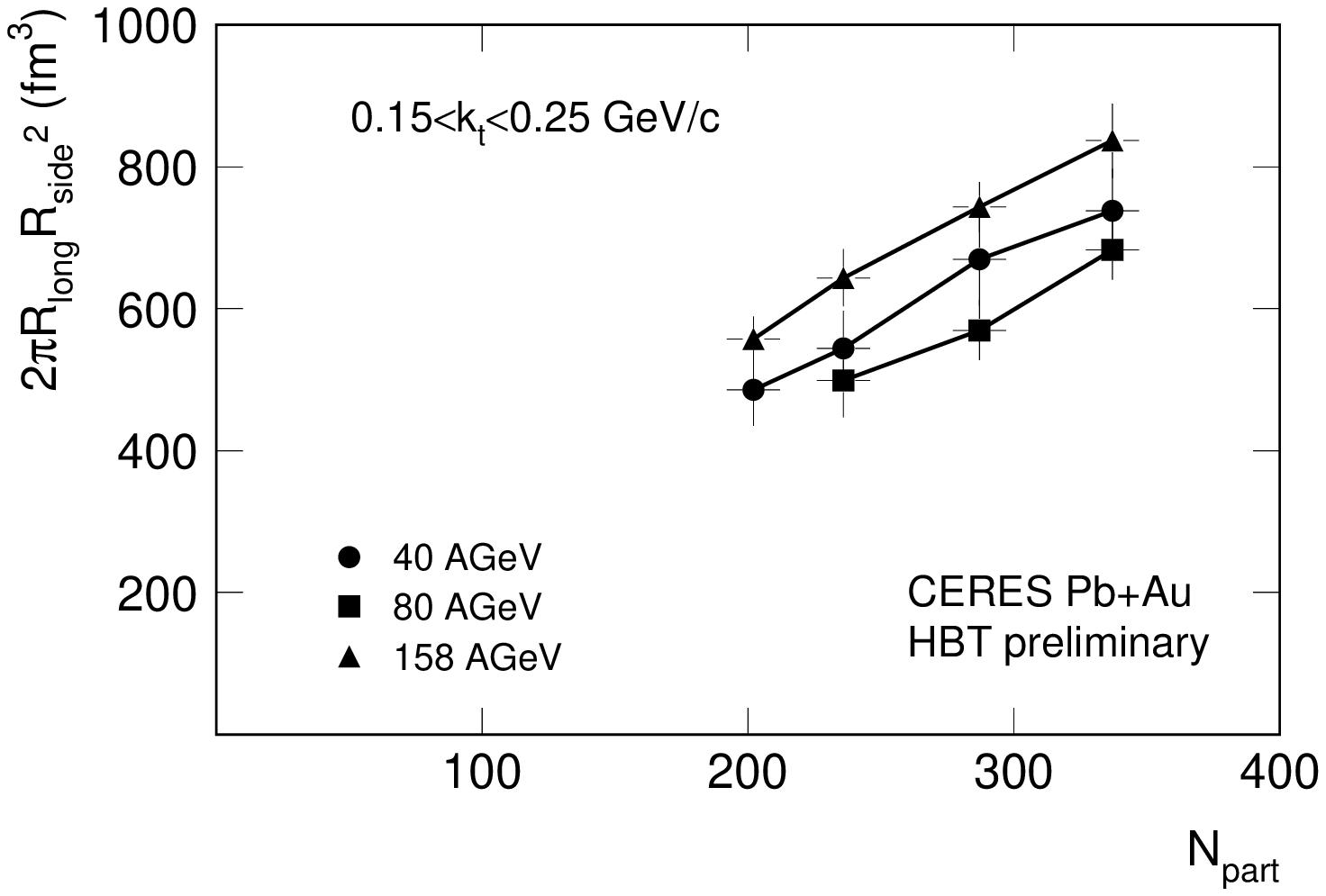}\\
   \parbox{14cm}
        {\centerline {\footnotesize 
        Fig.~5: $V_{f}$ as function of $N_{\rm part}$ at different SPS energies~\cite{heinz}.}}
\end{center}

\vspace{-0.5cm}

\vspace{-0.5cm}

\begin{center}
   \includegraphics[width=15cm]{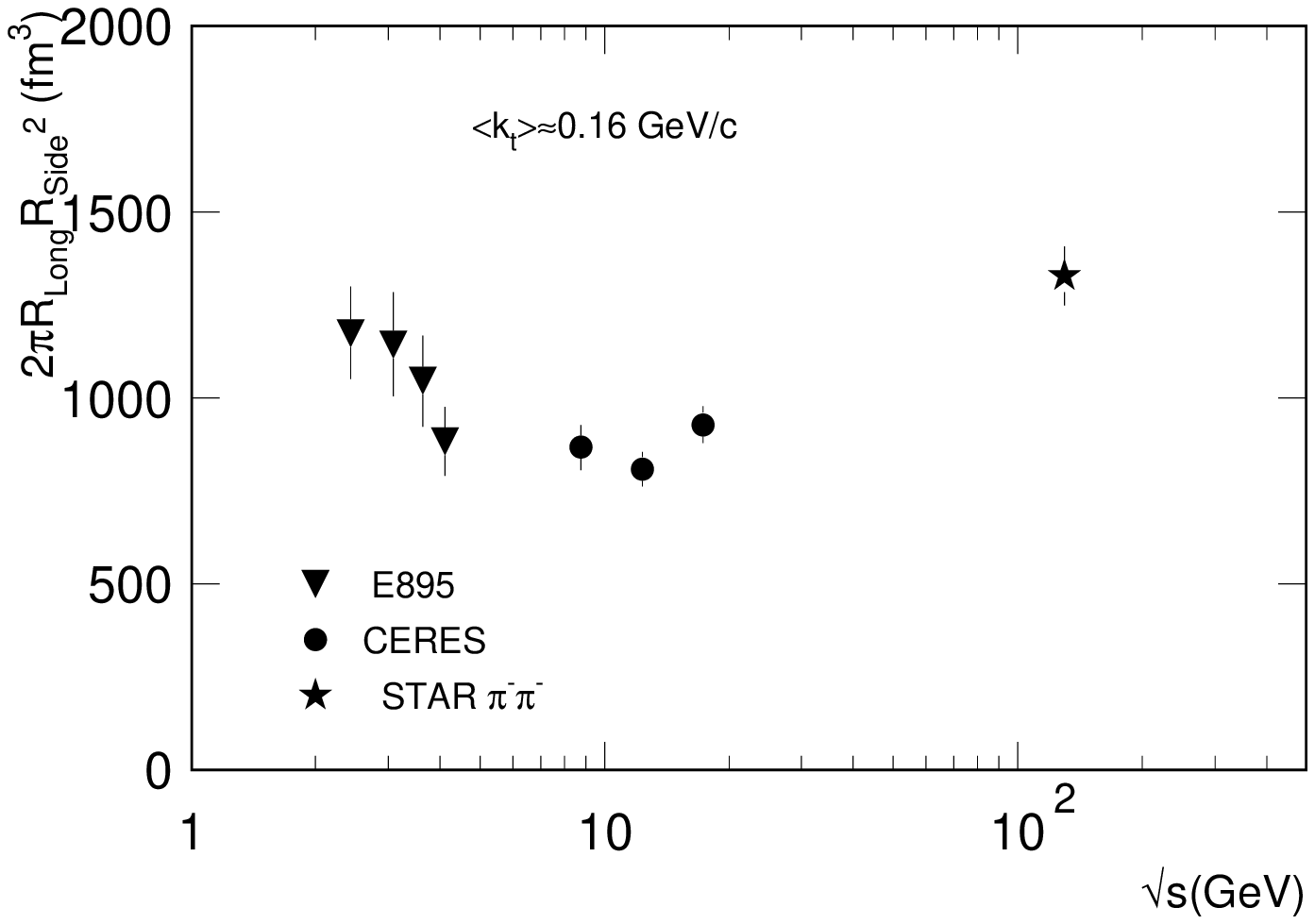}\\
   \parbox{14cm}
        {\footnotesize 
        Fig.~6: $V_{f}$ in central Pb(Au)+Pb(Au) collisions as function of $\sqrt{s}$ 
	(see text).  E895 points are at $<k_{t}>$=0.15~GeV/c~\cite{e895hbt}, 
	STAR is at $<k_{t}>$=0.17~GeV/c~\cite{starhbt}, CERES points are the average of 
	their bins at $<k_{t}>$=0.12 and $<k_{t}>$=0.2~GeV/c~\cite{heinz}.}
	\label{fig:vol_edep}
\end{center}

But the beam energy dependence is surprising: There is no clear hierarchy 
visible in Fig.~5
as expected from the increase of multiplicity by about 50\%
between 40~AGeV and 158~AGeV;
the smallest $V_{f}$ are observed at 80 AGeV. Obviously, the freeze-out
volume scales with multiplicity as long as multiplicity is controlled via
centrality, but it does not scale accordingly as multiplicity changes with
beam energy.

The comparison of $V_{f}$ at different beam energies from AGS to RHIC
sheds some light on this: The freeze-out volume $V_{f}$ is gradually decreasing
within the AGS energy range, reaches a minimum at SPS and then increases towards
RHIC, as demonstrated in Fig.~6.

Clearly, a simple relation between multiplicity and freeze-out volume cannot
hold. On the other hand, it is plausible to assume that at low energies
pions interact mainly with nucleons, while at high energies pion-pion 
scattering dominates. Is the transition from nucleon to pion dominated
freeze-out characterized by a minimum in $V_{f}$ at SPS?

\vspace{-0.5cm}

\begin{center}
   \includegraphics[width=15cm]{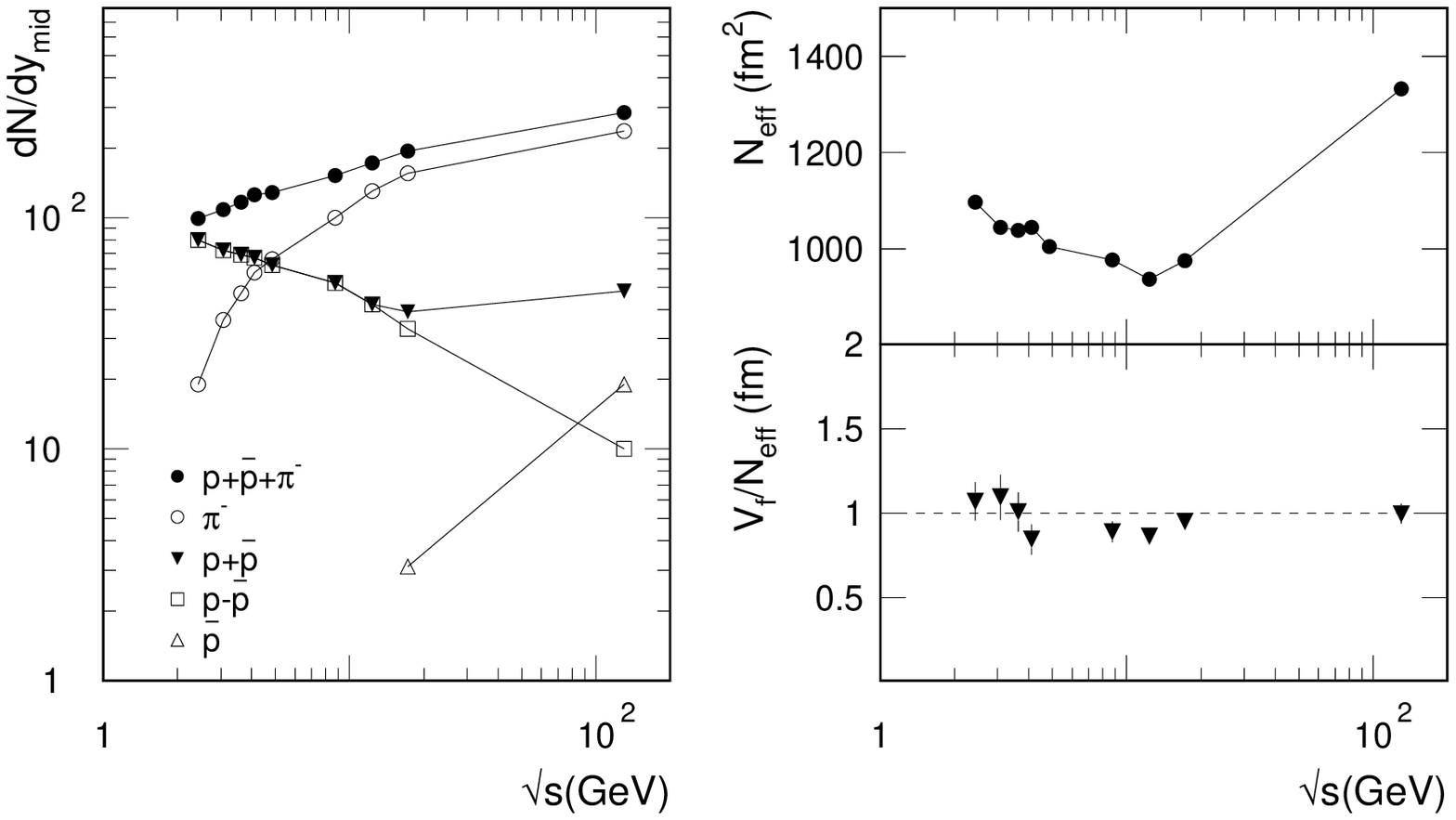}\\
   \parbox{14cm}
        {\footnotesize 
        Fig.~7: Left: Midrapidity yields of different particles vs.~$\sqrt{s}$. Upper 
	right: cross section weighted sum of particle yields (see text). Lower right:
	ratio $V_{f}/N_{\rm eff}$ vs.~$\sqrt{s}$.}
	\label{fig:dndy}
\end{center}

\vspace{-0.5cm}

In Fig.~7~(left) the midrapidity density of 
pions~\cite{cebra,seyboth,starmult,starqm} and
protons~\cite{e895prot,e917prot,ceresqm,na49prot,na49aprot,staraprot} in central
Pb(Au)+Pb(Au) collisions
is shown as function of $\sqrt{s}$. The pion yield increases monotonically
with beam energy. The {\em total} proton (p+$\bar{\rm p}$) yield at midrapidity drops from AGS
to SPS and stays approximately constant between SPS and RHIC because the decreasing number
of net protons is compensated by p-$\bar{\rm p}$ production. The sum of pions and
protons, however, is still a monotonic function of $\sqrt{s}$. 

At this point, the different cross sections $\sigma_{\pi\pi}$ and $\sigma_{\pi N}$ have 
to be considered. In Fig.~7~(upper right) the cross section
weighted sum of pions and protons
$N_{\rm eff}$=$2\cdot N_{{\rm p}+\bar{\rm p}} \cdot \sigma_{\pi N}$+
$3\cdot N_{\pi^{-}} \cdot \sigma_{\pi \pi}$ is shown \footnote{Note that
there are a few simplifications: the protons are multiplied by 2 to account
for the neutrons, which is not completely correct for the lowest
beam energies. Also the role of light nuclei and other produced particles
is neglected.}. 
For the cross sections
$\sigma_{\pi\pi}$=10~mb and $\sigma_{\pi N}$=65~mb are assumed.
Indeed, the cross section weighted sum of pions and nucleons at midrapidity
is {\em non}-monotonic and exhibits a minimum at SPS. As a consequence, the 
ratio $V_{f}/N_{\rm eff}$, which has the dimension of a length, is 
approximately constant (Fig.~7,~lower right).
This result suggests $V_{f}/N_{\rm eff}$$\approx$1~fm as a universal freeze-out
condition. 

In this picture, the relatively weak beam energy dependence of HBT source
parameters can be understood as an interplay between the decreasing 
(and eventually levelling off) (p+$\bar{\rm p}$) yield
and the increasing pion multiplicity at midrapidity, if their different
cross sections with pions are taken into account. It is interesting to obtain a detailed
understanding of the role of protons for pion freeze-out at low
energies. The strong momentum dependence of the $\sigma_{\pi N}$ cross section
may affect the $k_{t}$-dependence of $R_{\rm side}$ at low beam energies 
and cause the flattening of the $k_{t}$-dependence with increasing beam energy,
as the importance of protons ceases. At the lower SPS energies, where
protons are still important, the interpretation of the $k_{t}$-dependence
of $R_{\rm side}$ in terms of radial flow may be questionable, and
possibly breaks down at AGS.

\section{Discussion}
Single particle spectra and azimuthal anisotropies at SPS and RHIC have
been well reproduced by state-of-the-art hydro+cascade 
calculations~\cite{soff,soffhirsch,teaneyqm}.
However, they fail to reproduce the measured HBT parameters,
in particular the lifetime of the system is grossly overestimated.

In these models, the early plasma phase is described by hydrodynamics assuming  
an ideal plasma EOS with a first order phase transition and a mixed 
phase with an adjustable latent heat. At the end of the mixed phase,
hadrons are created assuming a thermal phase space population superimposed by
the collective velocity field created by the plasma pressure.
After hadronization, hadrons are rescattered using a hadronic cascade code.

In \cite{teaney} the response of the hydro+cascade calculation to different 
choices of the latent heat is well documented. From this investigation,
it becomes obvious that the choice of a relatively large latent heat of
0.8~GeV/fm$^3$, needed to describe the SPS single particle data, is
mainly driven by the observed mass dependence of the inverse slope parameters.
Smaller latent heat produces larger flow velocities {\em before} hadronization
and therefore too hard spectra, in particular for the multi-strange baryons
which do not follow the linear scaling behaviour observed for $\pi$, K, p~\cite{xu}.
On the other hand, a small latent heat would lead to an early acceleration 
of the matter and therefore to a faster dilution and shorter lifetime, 
in particular of the hadronic phase, which is suggested by the HBT measurements.

This leads to a closer inspection of the data as represented 
in Fig.~8 (left). 
For some of the particle species, the observed inverse slope parameter
scales linearly with the particle mass, consistent with the assumption
of a common flow velocity. There are, however, a number of exceptions
which do not follow this rule ($\phi$, $\Xi$, $\Omega$). This was explained by 
their small cross sections with the surrounding matter, hence they are
expected to participate less in the collective motion.

\vspace{-0.5cm}

\begin{center}
   \includegraphics[width=15cm]{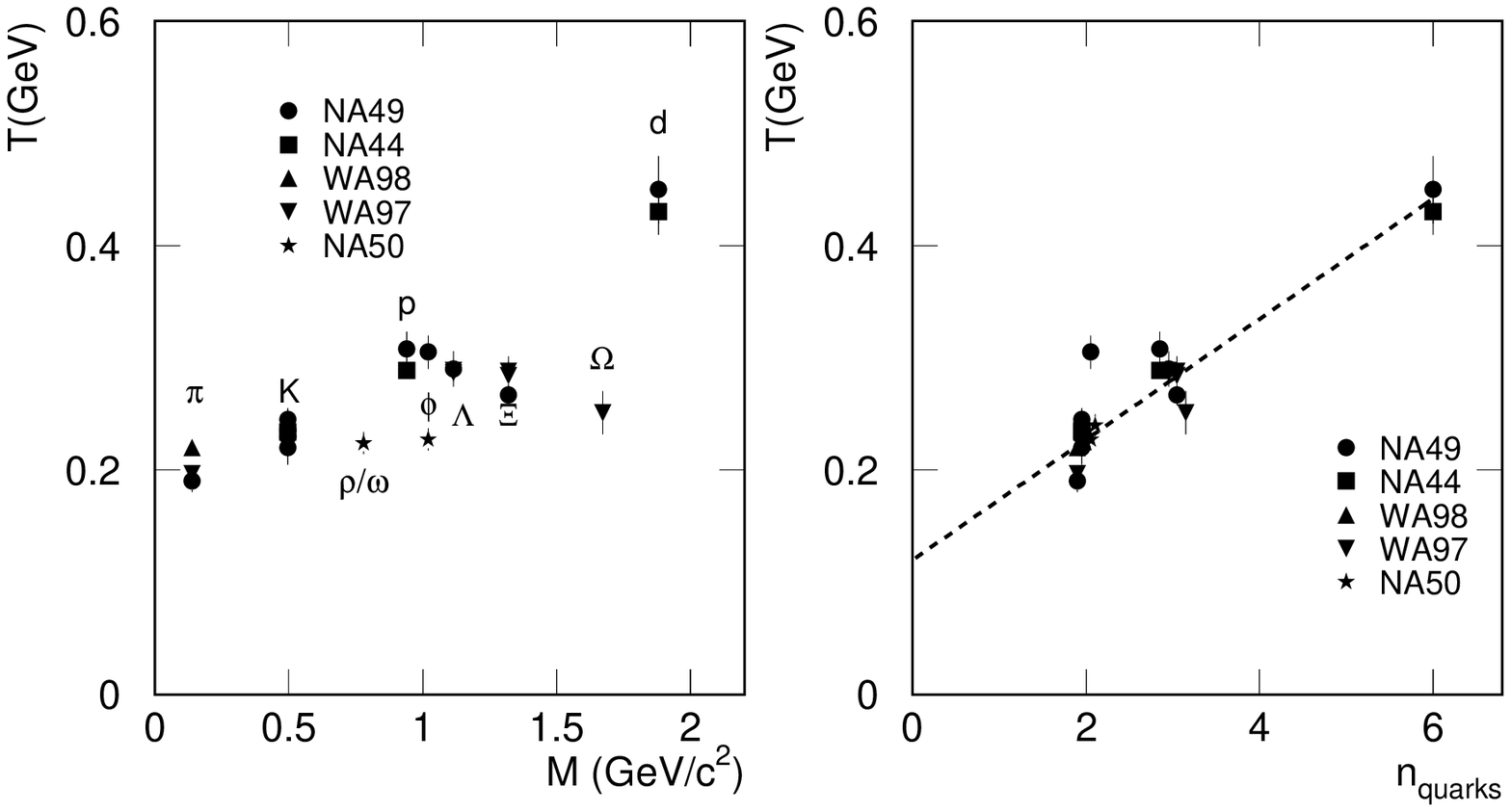}\\
   \parbox{14cm}
	{\footnotesize 
        Fig.~8: Left: Inverse slope parameter vs.~mass of hadrons at SPS. Right: 
	Inverse slope parameter vs.~number of constituent quarks.}
	\label{fig:mdep}
\end{center}

\vspace{-0.5cm}

On the other hand, it is evident that the data separate into two groups: 
All mesons have similar
slopes, and the same is true for baryons. One may therefore as well correlate 
the inverse slope with the number of constituent quarks inside the
hadron, rather than with the hadron mass (Fig.~8, right).
Also here a scaling behaviour can be observed, and almost all particles
are consistent with this trend.
Note that the large inverse slope
of the $\phi$ observed by NA49~\cite{na49phi} is 
experimentally still under debate, since
there is also a measurement by NA50~\cite{na50phi}, giving a much smaller number. 

The physical picture which arises from this representation is different
from the common interpretation: collective motion has completely
developed {\em before} hadronization, with constituent quarks being the flowing
objects. Hadrons are formed by coalescence of constituent quarks, preferentially
if they are close in phase space, thereby adding their momenta.
When hadrons acquire mass, momentum is
conserved which leaves the momentum spectra unchanged. 
In this picture, the data are consistent with a kinetic freeze-out temperature
of 0.12~GeV, a collective quark flow velocity of $\sqrt{1/3}$ and an effective
quark mass of 0.33~GeV: $T$=$0.12+\frac{1}{2} \frac{1}{3}\cdot 0.33 \cdot n_{\rm quarks}$
(dashed line in Fig.~8, right). 
Most of the explosive power of the system 
would be assigned to the early, pre-hadronic phase of the
collision, thereby qualitatively explaining the 
short lifetimes observed by HBT. 
It also suggests that the hadronic mass scale is 
no more relevant in the early phase of the collision, as
naively expected for a deconfined partonic system.


\end{document}